\documentclass[pdftex,12pt]{article}
\usepackage{amsmath,amsfonts,amsbsy,latexsym,amssymb,amscd,amstext}
\usepackage{pst-node}
\usepackage{dsfont}
\usepackage{cool}
\usepackage{overpic,youngtab}
\pdfoutput=1
\usepackage{rotating}
\usepackage{amsmath}
\usepackage{amssymb}
\usepackage{amsthm}
\usepackage{tensor}
\usepackage{tikz}
\usepackage{slashed}
\usepackage{commath}
\usepackage{cleveref}
\usepackage{graphicx}
\usepackage{caption}
\usepackage{subcaption}
\def\be{\begin{equation}}
\def\ee{\end{equation}}
\def\bea{\begin{eqnarray}}
\def\eea{\end{eqnarray}}
\def\bpm{\begin{pmatrix}}
	\def\epm{\end{pmatrix}}
\def\be{\begin{equation}}

\def\ee{\end{equation}}
\def\bea{\begin{eqnarray}}

\def\eea{\end{eqnarray}}

\def\a{\alpha}

\def\b{\beta}

\def\g{\gamma}
\def\d{\delta}

\def\m{\mu}

\def\n{\nu}

\def\l{\lambda}

\def\r{\rho}

\def\s{\sigma}

\def\bma{\begin{pmatrix}}
	\def\ema{\end{pmatrix}}

\def\bi{\begin{itemize}}
	\def\ei{\end{itemize}}

\begin{document}

		\vspace*{-1cm}
		\phantom{hep-ph/***} 
		{\flushleft
			{{FTUAM-xx-xx}}
			\hfill{{ IFT-UAM/CSIC-22-68}}}
		\vskip 1.5cm
		\begin{center}
		{\LARGE\bfseries  The fate  of horizons  under quantum corrections.}\\[3mm]
			\vskip .3cm
		
		\end{center}

		\vskip 0.5  cm
		\begin{center}
			{\large Enrique \'Alvarez, Jes\'us Anero and Eduardo Velasco-Aja.}
			\\
			\vskip .7cm
			{
				Departamento de F\'isica Te\'orica and Instituto de F\'{\i}sica Te\'orica, 
				IFT-UAM/CSIC,\\
				Universidad Aut\'onoma de Madrid, Cantoblanco, 28049, Madrid, Spain\\
				\vskip .1cm

				\vskip .5cm
				
				\begin{minipage}[l]{.9\textwidth}
					\begin{center} 
							\textit{E-mail:} 
						\tt{enrique.alvarez@uam.es},
						\tt{jesusanero@gmail.com} and
						\tt{eduardovelascoaja@gmail.com}
					\end{center}
				\end{minipage}
			}
		\end{center}
	\thispagestyle{empty}
	
\begin{abstract}
	\noindent
	In this paper, we study the physical effects of deforming the usual
	Einstein-Hilbert lagrangian with the Goroff-Sagnotti counterterm (the
	first which is nonvanishing on shell). Only spherically symmetric solutions are discussed. The facts that Schwarzschild’s spacetime is not a solution to the corresponding equations of motion
	and Birkhoff’s theorem validity are analyzed and some
	consequences worked out.
\end{abstract}

\newpage
\tableofcontents
	\thispagestyle{empty}
\flushbottom

\newpage
%%%%%%%%%%%%%%%%%%%%%%%%%%%%%%%%%%%%%%%%%%%%%%%%%%%%%%%%%%%%%%%%%%%%%%%%%%%%%%%%%%%%%%%%%%%%%%%%%%%%%%%%%%%%%%%%%%%%%%%%%%%%%%
\section{Introduction. }
%%%%%%%%%%%%%%%%%%%%%%%%%%%%%%%%%%%%%%%%%%%%%%%%%%%%%%%%%%%%%%%%%%%%%%%%%%%%%%%%%%%%%%%%%%%%%%%%%%%%%%%%%%%%%%%%%%%%%%%%
From the vantage point of effective theories, it seems unavoidable to adopt the point of view that the Einstein-Hilbert lagrangian (linear in curvature) is but the lowest energy approximation (deep infrared) of a more general lagrangian involving higher powers of curvature (maybe all of them). It is then important to reappraise the standard lore of gravitational collapse and examine to what extent it is stable under perturbations by higher dimensional operators in the gravitational lagrangian. For lack of a better name, we shall denote this endeavor with the slightly pretentious name of {\em structural stability}.
The basic definition of structural stability  (\cite{Arnold}, p, 92) by mathematicians is as follows. An ordinary differential equation (ODE) system 
\be
\dot{x}\equiv {dx\over dt}=\xi(x),
\ee
where $x\in \mathbb{R}^n$, is structurally stable when it remains equivalent to itself after a small variation of the vector field
\be
\xi\rightarrow \xi+\d\xi.
\ee

\par
Our main purpose is to examine the effect of changing the equations of motion as a result of the presence in the gravitational lagrangian of higher-order terms; specifically, quadratic and cubic in curvature. In fact, given that the on-shell counterterm \cite{tHooft} vanishes at one-loop,
\be
L^{(1)}_\infty= {1\over n-4}{1\over (4\pi)^2 }\int d^n x\sqrt{|g|} \left({1\over 60}\, R^2+ {7\over 10}\,R_{\m\n}^2\right).\label{Linf1} 
\ee
It is then essential to consider also the first non-vanishing on shell counterterm which appears at two-loop order, the Goroff-Sagnotti (GS) operator \cite{Goroff} and is proportional to the cube of Weyl's tensor.
\be
L_\infty^{(2)}={1\over n-4}{209\over 2880}{1\over (4\pi)^4 }\int d^n x\sqrt{|g|}  \, W_{\m\n\a\b}W^{\a\b\r\s}W_{\r\s}^{~~\m\n},\label{Linf2}
\ee
note that for $n \leq 5$ there is only one independent such contraction, \cite{Fulling}.
\par
The generic lagrangian for the gravitational field, from this point of view, would be
\be
L=\sqrt{|g|}\bigg\{-\frac{1}{2\kappa^2} R+ \omega\kappa^2 \, W_{\m\n\a\b}W^{\a\b\r\s}W_{\r\s}^{~~\m\n}\bigg\},\label{Action}
\ee
with dimensionless coupling constant $\omega$. 

Let us remind the reader that  Weyl's tensor is defined as the totally traceless piece of Riemann's tensor, and depends explicitly on the spacetime dimension, $n$:
\bea
&W_{\a\b\g\d}\equiv R_{\a\b\g\d}-{1\over n-2}\bigg\{g_{\a\g} R_{\b\d}-g_{\a\d} R_{\b\g}-g_{\b\g}R_{\d\a}+g_{\b\d} R_{\g\a}\bigg\}+\nonumber\\
&+{1\over (n-1)(n-2)}\,R\,\bigg\{g_{\a\g} g_{\d\b}-g_{\a\d} g_{\g\b}\bigg\}.
\eea

\par
What we mean is the following. The coefficients of the quadratic operators $R^2$ and $R_{\m\n}^2$ in \eqref{Linf1} can be put equal to zero without fine tuning because quantum corrections will not disrupt this. This is not the case with the coefficient in front of the Goroff-Sagnotti (GS) counterterm, which would be in any case generated by quantum corrections even if it was initially set to zero. Is in this sense that the lagrangian \eqref{Action} is the simplest one incorporating quantum corrections.

\par
Ideally, we would want to get exact solutions to the new equations, but those are few and far between. Therefore, to obtain information about quantum corrections, we will rely on approximation methods: perturbation theory and some modification of Frobenius' power series method.
\par
In this paper, we will consider only spherically symmetric perturbations\footnote{Hence, when talking about {\em solutions} we will mean spherically symmetric solutions.}. It is obvious that any trend of instability will persist in the general situation; whereas stability would have to be confirmed with a more general analysis.

\par
The plan of the paper is as follows.

In section \ref{sec:2}, we first present a necessary and sufficient condition to obtain solutions to the equations of motion (EM) corresponding to the GS counterterm. It turns out that such condition contains the solution to any\footnote{As we will discuss, the only exception to this is conformal gravity for which additional solutions exist.} four-dimensional action corresponding to the contraction of \(p\) Weyl tensors.
\par
This will also allow us to review, in a broad sense, the fate of Birkhoff's theorem when higher dimensional operators are taken into account. A more ambitious approach would be to study in detail models of gravitational collapse \cite{Christodoulou}; we leave that for a future task.
In the context of theories of gravity with lagrangians involving higher powers of the Riemann tensor, Birkhoff's theorem has been proved for Lovelock theories \cite{Deser}  as well as for conformal gravity \cite{Riegert}, but it does not hold for pure Riemann squared lagrangians \cite{Kehm}.
\par

The rest of the paper is dedicated to getting more information on solutions of the complete lagrangian \eqref{Action}. For this in section \ref{sec:3}, we resort to a formal power series analysis to obtain solutions both near the origin and far from the origin. Such an analysis shows the presence of a singular \((2,2\) solution similar to the one discussed extensively in \cite{Holdom} for quadratic gravity. Additionally, we also report in more detail the existence of regular \((0,0)\) solutions.

Lastly, in section \ref{sec:4} we use a perturbative analysis to study the structural stability of Schwarzschild's spacetime (SST) subject to the GS deformation. It seems that the GS perturbation may be able to remove the Event Horizon.

%%%%%%%%%%%%%%%%%%%%%%%%%%%%%%%%%%%%%%%%%%%%%%%%%%%%%%%%%%%%%%%%%%%%%%%%%%%%%%%%%%%%%%%%%%%%%%%%%%%%%%%%%%%%%%%%%%%%%%%%%%%%%%
\section{On the solution to \(W^p\) actions.}\label{sec:2}
%%%%%%%%%%%%%%%%%%%%%%%%%%%%%%%%%%%%%%%%%%%%%%%%%%%%%%%%%%%%%%%%%%%%%%%%%%%%%%%%%%%%%%%%%%%%%%%%%%%%%%%%%%%%%%%%%%%%%%%%%%%%%%
%It is well known that Weyl's tensor is conformally invariant. Under
%\be
%\ee
In section \ref{sec:2.1}, we present a necessary and sufficient condition for a four-dimensional spherically symmetric spacetime to be a solution to the EM, whenever \(p\neq0\),  the conformal dimension. 

We then go into more detail about how this condition fails to capture all solutions for conformal gravity (section \ref{sec:2.3}), i.e. \(p=2\). This discussion serves as an indication of the fact that general \(W^p\) actions satisfy Birkhoff's Theorem.	

After this, in section \ref{sec:2.2}, we will use this general observation for the specific \(p=3\) case to provide some examples of solutions to the GS EM.

Lastly, section \ref{sec:2.4} explores how this property allows one to read some conditions on a solution for the full action \eqref{Action}.
%%%%%%%%%%%%%%%%%%%%%%%%%%%%%%%%%%%%%%
\subsection{A necessary and sufficient condition for solutions to \(W^{p\neq2}\) actions.}\label{sec:2.1}
The metric for a spherically-symmetric spacetime can be written \cite{Carroll2019} as,
\begin{equation}
ds^2= B(r,t) dt^2-A(r,t) dr^2-r^2 d\Omega_2^2.\label{metric2} 
\end{equation}
Applying the Principle of Symmetric criticality (PSC) \cite{Weyl, Palais, Deser}, we can obtain solutions varying the restricted group-invariant Lagrangian\footnote{Which is calculated using metric \eqref{metric2}.} with respect to the fields \(A(r,t),\, B(r,t)\). 
The EM will then correspond to the two Euler-Lagrange equations for the fields \(A(r,t),\, B(r,t)\).

Omitting the explicit \((r,t)\) dependence, the restricted group-invariant lagrangian for the GS counterterm corresponds to,
 \begin{equation}
 S_{\tiny{Weyl^3}}=\omega\kappa^2\int dt\,dr\;r^2\sqrt{AB} \Bigg[\frac{G(A,A',\dot{A},\ddot{A},B,B',B'',\dot{B}) }{\sqrt[3]{144} r^2 A^2 B^2}\Bigg]^3.\label{ActionW3PSC}
 \end{equation}
 With  $\dot{A}=\partial_t A(r,t)$ and $A'=\partial_r A(r,t)$ and \(G\) equal to,
\begin{align}
    G(A,A',B,B',\cdots)&:=A \left(r^2 \left(B'^2-\dot{A} \dot{B}\right)+2 r B \left(r \ddot{A}+B'-r B''\right)-4 B^2\right)+\nonumber\\
    &+r B \left(A' \left(r B'-2 B\right)-r \dot{A}^2\right)+4 A^2 B^2.
 \end{align}
The fact that \(p\) contractions of a Weyl tensor have such a simple structure was first noted for static and spherically symmetric spaces by Deser and Ryzhov in \cite{Deser}.

Here we show that this holds even when the staticity condition is lifted. This can be seen by noting that Weyl's tensor can be expressed as 
\be
W_{\m\n\r\s}=G\left[A,A^\prime,\dot{A},\ddot{A},B,B^\prime,\dot{B},B^{\prime\prime},r\right]\, C_{\m\n\r\s},
\ee
and $C_{\m\n\r\s}$ is a tensor depending on the metric components but not on its derivatives. 
\\
In detail
\bea
C_{01\a\b}=-C_{10\a\b}&&=-\frac{1}{12}\begin{pmatrix}
   0&-g^{22}&0&0\\
   g^{22}&0&0&0\\
   0&0&0&0\\
   0&0&0&0\end{pmatrix}g^{00}g^{11}\nonumber\\
C_{02\a\b}=-C_{20\a\b}&&=-\frac{1}{12}\begin{pmatrix}
   0&0&\frac{1}{2}g^{11}&0\\
   0&0&0&0\\
   -\frac{1}{2}g^{11}&0&0&0\\
   0&0&0&0\end{pmatrix}g^{00}g^{11}\nonumber\\
C_{03\a\b}=-C_{30\a\b}&&=-\frac{1}{12}\begin{pmatrix}
   0&0&0&\frac{1}{2}g^{11}g^{22}g_{33}\\
   0&0&0&0\\
   0&0&0&0\\
   -\frac{1}{2}g^{11}g^{22}g_{33}&0&0&0\end{pmatrix}g^{00}g^{11}\nonumber\\
C_{12\a\b}=-C_{21\a\b}&&=-\frac{1}{12}\begin{pmatrix}
   0&0&0&0\\
   0&0&\frac{1}{2}g^{00}&0\\
   0&-\frac{1}{2}g^{00}&0&0\\
   0&0&0&0\end{pmatrix}g^{00}g^{11}\nonumber\\
C_{13\a\b}=-C_{31\a\b}&&=-\frac{1}{12}\begin{pmatrix}
   0&0&0&0\\
   0&0&0&\frac{1}{2}g^{00}g^{22}g_{33}\\
   0&0&0&0\\
   0&-\frac{1}{2}g^{00}g^{22}g_{33}&0&0\end{pmatrix}g^{00}g^{11}\nonumber\\
C_{23\a\b}=-C_{32\a\b}&&=-\frac{1}{12}\begin{pmatrix}
   0&0&0&0\\
   0&0&0&0\\
   0&0&0&-g^{00}g^{11}g_{33}\\
   0&0&g^{00}g^{11}g_{33}&0\end{pmatrix}g^{00}g^{11}
\eea
%%%%%%%%%%%%%%%%%%%%
These properties are common to all lagrangians built out of powers of the Weyl tensor. 

We now use this observation to prove the following.

\textbf{Claim:} For a spherically symmetric action consisting of \(p\) contracted Weyl tensors condition,
\begin{equation}
    G\left[A,A^\prime,\dot{A},\ddot{A},B,B^\prime,\dot{B},B^{\prime\prime},r\right]=0,\label{G1}
\end{equation}
contains all non-singular solutions to the EM.

\textbf{Proof:}

Let us generalize a little bit the GS counterterm and consider the general  action in $n=4$ dimensions
\be
S_{\tiny{Weyl^p}}\equiv \int d^4x \sqrt{|g|} W_{\m_1\n_1 \r_1\s_1 }\ldots W_{\m_p\n_p\r_p\s_p}\,I^{{\vec\m}{\vec\n}{\vec\r}{\vec\s}},\label{Weylp}
\ee
with the tensor constructed out of metric tensors
\be
I^{{\vec\m}{\vec\n}{\vec\r}{\vec\s}}\equiv  g^{\m_{i_1}\n_{j_1}}\ldots g^{\r_{k_p} \s_{l_p}} \quad (\text {$2p$ terms}),
\ee
where $(i_1\ldots i_p)$, $(j_1\ldots j_p)$, $(k_1\ldots k_p)$, $(l_1\ldots l_p)$ are permutations of $(1\ldots p)$, represents all possible scalars containing $p$ Weyl tensors. For example, the GS operator corresponds to
\be 
I^{{\vec\m}{\vec\n}{\vec\r}{\vec\s}}=g^{\m_1 \r_3} g^{\n_1 \s_3} g^{\r_1 \m_2} g^{\s_1\n_2} g^{\r_2 \m_3} g^{\s_2 \n_3}.
\ee

The EM can be formally expressed as 
\be \frac{1}{2}g_{\a\b}G^pC_{\m\n\r\s}^p+pG^{p-1}\frac{\d G}{\d g_{\a\b}}C_{\m\n\r\s}^p+G^p\frac{\d C_{\m\n\r\s}^p}{\d g_{\a\b}}=0,\ee
where
\be C_{\m\n\r\s}^p= C_{\m_1\n_1 \r_1\s_1 }\ldots C_{\m_p\n_p\r_p\s_p}\,I^{{\vec\m}{\vec\n}{\vec\r}{\vec\s}}.\ee

This means that all solutions of  the ODE
\be
G\left[A,A^\prime,\dot{A},\ddot{A},B,B^\prime,\dot{B},B^{\prime\prime},r\right]=0,\nonumber
\ee
are also solutions of the EM ({\it sufficient condition}). They depend on an arbitrary function.

\par
Also, since the action is conformal in  dimension $n=2p$, the trace of the EM is proportional to
\be
g^{\a\b}{\d S_{\tiny{Weyl^p}}\over \d g_{\a\b}}\propto (2p-4) W_{\m_1\n_1 \r_1\s_1 }\ldots W_{\m_p\n_p\r_p\s_p}\,I^{{\vec\m}{\vec\n}{\vec\r}{\vec\s}}.
\ee
This is forced by conformal invariance and dimensional arguments. Indeed,   Deser-Schwimmer's \cite{DeserS} conjecture, proved by Alexakis \cite{Alexakis}, states that the most general form of the integrand of a conformal (Weyl) invariant is of the form
\be
P\equiv W+\nabla_\m j^\m+ c.\text{Pf}\left(R_{\m\n\r\s}\right),
\ee
where $W$ is a conformal invariant (i.e. $\sqrt{|g|} W^2$ in n=4 dimensions), and the pfaffian is proportional to the integrand of Euler's characteristic of a compact manifold $M$ through
\be
\int_M\,\text{Pf}\left(R_{\m\n\r\s}\right)\,d(vol)={2^n \pi^{n\over 2} \left(n/2 -1\right)!\over 2 (n-1)!}\chi(M).
\ee
Finally $c$ is a constant.

The fact that $G=0$ are the only solutions to $Weyl=0$ (barring divergences in the metric tensor) means that for all  powers $p\neq 2$ we capture all solutions ({\it necessary condition}) of the EM by simply solving 
\be
G\left[A,A^\prime,\dot{A},\ddot{A},B,B^\prime,\dot{B},B^{\prime\prime},r\right]=0.\nonumber
\ee
We stress this is no longer true in the conformal dimension (that is $p=2$). This is the reason why SST is a viable solution for four-dimensional conformal gravity \cite{Mannheim}.
%%%%%%%%%%%%%%%%%%%%%%%%%%%%%%%%%%%%%%%%%%%%%%%%%%%%%%%%%%%%%%%%%%%%%%%%%%%%%%%%%%%%%%%%%%%%%%%%%%%%%%%%%%%%%%%%%%%%%%%%%%%%%%
\subsection{The exceptional case: Conformal Gravity \(p=2\).}\label{sec:2.3}
%%%%%%%%%%%%%%%%%%%%%%%%%%%%%%%%%%%%%%%%%%%%%%%%%%%%%%%%%%%%%%%%%%%%%%%%%%%%%%%%%%%%%%%%%%%%%%%%%%%%%%%%%%%%%%%%%%%%%%%%%%%%%%

It is well known that Weyl's tensor is conformally invariant. Under
\be
g_{\m\n}\rightarrow \Omega(x)^2 g_{\m\n},\label{conformal}
\ee
it transforms as 
\be
W^\m\,_{\n\r\s}\rightarrow \, W^\m\,_{\n\r\s}.
\ee
The Weyl squared one is conformally (gauge) invariant in n=4 dimensions
\be
S_{\tiny{Weyl^2}}=\int d^4 x \sqrt{|g|} \, W_{\m\n\a\b}W^{\m\n\a\b}.\label{actionW3}
\ee
The equations of motion \cite{AlvarezGM} imply the vanishing of the traceless Bach's tensor
\be
H_{\a\b}\equiv\frac{\d S}{\d g^{\a\b}}\equiv B_{\a\b}\equiv K^{\m\n} W_{\m\a\b\n} +\nabla^\l \left(\nabla_\l K_{\a\b}-\nabla_a K_{\b\l}\right),\label{Bach}
\ee
where Schouten's tensor \(K_{\m\n}\) is defined as
\be
K_{\a\b}\equiv {1\over n-2}\left(R_{\a\b}-{1\over 2(n-1)} R g_{\a\b}\right).
\ee
Physically the most important property of Bach's tensor is that it corresponds to a primary  operator of dimension 2; that is, under a Weyl rescaling \eqref{conformal} it transforms as
\be
B_{\m\n}\rightarrow \Omega^{-2} B_{\m\n}.
\ee

It was shown in \cite{Mannheim} that conformal gravity admits a solution,
\begin{equation}
    ds^2=\left(c_1+\frac{c_2}{r}+c_3 r+c_4r^2\right)dr^2-\frac{dt^2}{\left(c_1+\frac{c_2}{r}+c_3 r+c_4r^2\right)}-r^2d\Omega_2^2,
\end{equation}
while Deser and Tekin show \cite{Deser} that the \(GS\) counterterm does not allow for the \(c_2\) term. 
This is again because for the conformal dimension \(p=2\), the trace identically vanishes and therefore \(G=0\) will capture some but not all spherically symmetric and static solutions corresponding to, 
\begin{equation}
    \tensor{H}{_\a_\b}=0.
\end{equation}
This means that the set of all spherically symmetric solutions corresponding to \eqref{G1} belongs to the larger set of solutions of conformal gravity.

Taking into account the proof that conformal gravity satisfies Birkhoff's theorem \cite{Riegert}, i.e., any spherically-symmetric solution is static, and the fact that the set of solutions to conformal gravity contains the solutions for any action of the form \eqref{Weylp} we conclude that;

{\em Any spherically-symmetric solution to an action made of by \(p\) contractions of the Weyl tensor will be static.}

Therefore, Birkhoff's theorem holds for this larger set of actions.

Guided by this fact, in the remaining of the paper we will consider static and spherically symmetric spacetimes, i.e., we reduce \eqref{metric2}
to
\begin{equation}
ds^2= B(r) dt^2-A(r) dr^2-r^2 d\Omega_2^2.\label{metric} 
\end{equation}

%%%%%%%%%%%%%%%%%%%%%%%%%%%%%%%%%%%%%%%%%%%%%%%%%%%%%%%%%%%%%%%%%%%%%%%%%%%%%%%%%%%%%%%%%%%%%%%%%%%%%%%%%%%%%%%%%%%%%%%%%%%%%%
\subsection{Exact solutions for the Goroff-Sagnotti counterterm.}\label{sec:2.2}
%%%%%%%%%%%%%%%%%%%%%%%%%%%%%%%%%%%%%%%%%%%%%%%%%%%%%%%%%%%%%%%%%%%%%%%%%%%%%%%%%%%%%%%%%%%%%%%%%%%%%%%%%%%%%%%%%%%%%%%%%%%%%%

The general proof in section \ref{sec:2.1} is explicitly given\footnote{By considering the diff-invariant EM for the GS counterterm and their trace.} for the GS counterterm (\(p=3\)) in appendix \ref{sec:B}.

Let us now present some solutions to \eqref{EM} obtained by the condition proven in section \ref{sec:2.1}.
To do so we note that the condition \eqref{G1} is of first order in the derivatives of \(A(r)\) but of second order with respect to derivatives of \(B(r)\). Hence to find solutions we make an ansatz for \(B(r)\), (seed function), and integrate the first order ODE for \(A(r)\).
\begin{itemize}
   \item First we would like to recover the classical result, for the further simplification\footnote{This oversimplification works here even though the gauge  only depends on one function instead of two \cite{Deser}.} of \cite{Mannheim}, for that we further restrict the freedom of the pair \(A(r),\, B(r)\) and set
\begin{equation}
   B(r)=\frac{1}{A(r)}.
\end{equation}
Plugging this in \eqref{G1} yields, 
\begin{equation}
   \frac{-4 r^2 A'(r)^2+2 r A(r) \left(r A''(r)-2 A'(r)\right)+4 A(r)^3-4 A(r)^2}{A(r)^3}=0,
\end{equation}
which one can integrate to obtain,
\begin{equation}
   A(r)=\frac{1}{c_2 r^2+c_1 r+1}.
\end{equation}
Which corresponds to the solution first reported in \cite{Deser}. This further shows that there is no Schwarzschild solution as first pointed out by Deser and Tekin \cite{Deser}.
\end{itemize}
For the next two cases, we will consider independent \(A(r),\, B(r)\).
\begin{itemize}
\item We can obtain asymptotically flat solutions with Schwarzschild-like behavior at large \(r\) using the seed, 
\begin{equation}
B(r)=1-\frac{\mathcal{M}\,r^n}{r^{n+1}+\lambda},\qquad \mbox{With }\,\mathcal{M,\lambda}>0.\label{AFW3} 
\end{equation}
These solutions satisfy that, 
\begin{equation}
   \mbox{For }\quad r>>\sqrt[n+1]{\lambda}\qquad B(r)\simeq 1-\frac{\mathcal{M}}{r}.
\end{equation}
Some particular choices of \(n\) and corresponding \(A(r)\) are, 
\begin{itemize}
   \item \(n=0\)
   \begin{align}
	   B(r)&=1-\frac{\mathcal{M}}{r+\lambda},\\
	   A(r)&= \frac{\left(2 r^2+r (4 \lambda -3 \mathcal{M})+2 \lambda  (\lambda -\mathcal{M})\right)^2}{(\lambda +r)^2 (\lambda +r-\mathcal{M}) \left(c_1 r^2 (\lambda +r)+4 (\lambda +r-\mathcal{M})\right)}.
   \end{align}
   \item \(n=1\)
   \begin{align}
	   B(r)&=1-\frac{\mathcal{M}\,r}{r^{2}+\lambda},\\
	   A(r)&=\frac{\left(2 \lambda ^2+r^3 (2 r-3 \mathcal{M})+\lambda  r (4 r-\mathcal{M})\right)^2}{\left(\lambda +r^2\right)^2 \left(\lambda +r^2-r \mathcal{M}\right) \left(r \left(r \left(c_1 \left(\lambda +r^2\right)+4\right)-4 \mathcal{M}\right)+4 \lambda \right)}.
   \end{align}
   \item \(n=2\)
   \begin{align}
	   B(r)&=1-\frac{\mathcal{M}\,r^2}{r^{3}+\lambda},\\
	   A(r)&=\frac{\left(2 \lambda ^2+2 r^6-3 r^5 \mathcal{M}+4 \lambda  r^3\right)^2}{\left(\lambda +r^3\right)^2 \left(\lambda +r^3-r^2 \mathcal{M}\right) \left(r^2 \left(c_1 \left(\lambda +r^3\right)+4 r-4 \mathcal{M}\right)+4 \lambda \right)}.
   \end{align}
\end{itemize}
In all three cases above one needs to set \(c_1=0\) in \(A(r)\) to obtain regular at the origin and asymptotically flat solutions, i.e., such that for \(r> R\) for a certain scale, 
\begin{align}
   \mbox{If }\quad &r>>R, \\
   B(r)&\to1+\mathcal{O}(\frac{R}{r}),\\
   A(r)&\to1+\mathcal{O}(\frac{R}{r}).
\end{align}
\item \((2,2)\)-like solution. For its relevance to previous works and its presence in the power series analysis of section \eqref{sec:4}, we now provide an example of a solution to the GS counterterm that has a \((2,2)\) form at \(r\to0\).
\begin{equation}
   B(r)=r^2\mbox{exp}(-r).\label{B2}
\end{equation}
For this seed function, we recover, 
\begin{equation}
   A(r)=\frac{r^2}{c_1 \mbox{exp}(r)+4}.\label{A2}
\end{equation}
If we consider the expansions of equations \eqref{B2} and \eqref{A2}, we can see that they match that of the \((2,2)\) family in section \ref{sec:3},
\begin{align}
   B(r)&=r^2-r^3+\frac{r^4}{2}+\mathcal{O}\left(r^5\right),\\
   A(r)&=\frac{\left(c_1^2-4 c_1\right) r^4}{2 \left(c_1+4\right){}^3}-\frac{c_1 r^3}{\left(c_1+4\right){}^2}+\frac{r^2}{c_1+4}+\mathcal{O}\left(r^5\right).
\end{align}
Now one can check that the expansions above match exactly the one of equations \eqref{22.1} and \eqref{22.2}.
\end{itemize}
The unique condition \eqref{G1} allows for an infinite number of solutions most of which need not have a clear physical meaning. 
The solutions explored here are such that they resemble different features of the power series solutions obtained in section \eqref{sec:4}.

\subsection{The absence of Ricci flat solutions.}\label{sec:2.4}
%%%%%%%%%%%%%%%%%%%%%%%%%%%%%%%%%%%%%%%%%%%%%%%%%%%%%%%%%%%%%%%%%%%
The diff invariant EM for action \eqref{Action} are given by equation \eqref{EM}. 
Taking the trace,
\be
g^{\a\b}{\d S\over \d g^{\a\b}}\equiv  H^\l_\l=-\frac{(n-2)}{2\kappa^2}R+\omega\kappa^2(n-6)W^3=0.\ee

A conclusion one can extract from the trace of the EM is that \textbf{No Ricci flat nor conformally flat metric} can be a solution for the complete system. 
This can be seen as follows, 
\begin{itemize}
   \item If there were a conformally flat solution, which corresponds to, 
   \begin{equation}
   G=0\rightarrow\tensor{W}{^\a_\b_\gamma_\delta}=0\rightarrow W^3=0.
   \end{equation} 
   In this case, we would have that the trace equation is just, 
   \begin{equation}
   R=0.
   \end{equation}
   But these two conditions mean that the EM become, 
   \begin{equation}
   \tensor{R}{_\mu_\nu}=0,
   \end{equation}
   where for a static and spherically symmetric spacetime Birkhoff's theorem for GR holds and the only solution is a path of the SST. 
   This then leads to a contradiction since, as mentioned in \cite{Deser} and the series analysis performed in section \ref{sec:3}, SST is not a solution to the GS EM. Then we see \textbf{there are no conformally flat solutions} to the entire system.
   \item Then, we must have 
   \begin{equation}
   R=2\kappa^4 \omega \frac{(n-6)}{(n-2)}W^3\neq0,
   \end{equation}
   thus also preventing Ricci-flatness.
\end{itemize}
%%%%%%%%%%%%%%%%%%%%%%%%%%%%%%%%%%%%%%%%%%%%%%%%%%%%%%%%%%%%%%%%%%%%%%%%%%%%%%%%%%%%%%%%%%%%%%%%%%%%%%%%%%%%%%%%%%%%%%%%%%%%%%
\section{Some results of a  Frobenius-type analysis.}\label{sec:3}
%%%%%%%%%%%%%%%%%%%%%%%%%%%%%%%%%%%%%%%%%%%%%%%%%%%%%%%%%%%%%%%%%%%%%%%%%%%%%%%%%%%%%%%%%%%%%%%%%%%%%%%%%%%%%%%%%%%%%%%%%%%%%
While section \ref{sec:2} was dedicated to finding exact solutions to the EM of the GS counterterm, here we will try to find power series solutions to the \textbf{complete action}, 
\begin{equation}
   S=\int d^4x\left\{-\frac{1}{2\kappa^2}R+\omega\kappa^2W_{\m\n\a\b}W^{\a\b\r\s}W_{\r\s}^{~~\m\n}\right\}.
\end{equation}
In spite of the fact that Frobenius' theorem only holds for linear ODE, we shall seek solutions defined as power series, as previously done in \cite{Holdom, Stelle78, Lu, AASG}.
Owing to  spherical symmetry, made explicit in the gauge choice for the metric in \eqref{metric}, 
\begin{equation}
H_{\phi\phi}=\sin^2\left(\theta\right)H_{\theta\theta},\label{cond.1}
\end{equation}
and  Bianchi's identity, 
\begin{equation}
\nabla_\mu H^{\mu\nu}=0.\label{cond.2}
\end{equation}
only two out of the four non-vanishing components of $H_{\mu\nu}$  will remain independent.
\par
We will solve for the two independent ODEs corresponding to,
\begin{equation}
   \left\{
\begin{array}{rcl}
   H_{tt}&=0,\\
   H_{rr}&=0,
\end{array}
\right.\label{EMS}
\end{equation}
we include the full expressions for the EM \eqref{EMS} in appendix \ref{A}.

In section \ref{sec:3.1} we will consider series expansions near the origin of the form
\bea
&A(r) =a_s r^s+a_{s+1}r^{s+1}+\ldots,\nonumber\\
&B(r)= b_t(r^t+b_{t+1} r^{t+1}+\ldots),\label{metans}
\eea
and discuss the possible integer\footnote{As usual in such power series analysis \cite{Stelle78,Lu,AASG}.} values for \((s,t)\) as well as the properties of the curvature invariants \(R^2,\,\tensor{R}{_\mu_\nu^2},\,\tensor{R}{_\alpha_\beta_\gamma_\delta^2}\) at \(r=0\) where SST is known to have a curvature singularity.

In section \ref{sec:3.2} we will consider expansions for large \(r \) values corresponding to asimptotically flat spacetimes i.e.
\begin{align}
   A(r)&=1+\sum_n a_nr^{-n},\\
   B(r)&=1+\sum_n b_nr^{-n}.
\end{align}

From the obtained solutions we will see in which regions we will then extract some general features to base our discussion on the fate of horizons in section \ref{sec:4}.

\subsection{Power Series Solutions near the origin.}\label{sec:3.1}

The starting point of the procedure is to assume and expansion of the form
\bea
&A(r) =a_s r^s+a_{s+1}r^{s+1}+\ldots,\nonumber\\
&B(r)= b_t(r^t+b_{t+1} r^{t+1}+\ldots),
\eea
for integer \((s,t)\) values.
\subsubsection{Necessary conditions for \((s,t)\) families.}
We will separately study the case \((0,0)\) and \((s,t)\neq(0,0)\). 

In the latter case, there are terms in the expansion only proportional to the lowest order coefficients, \(a_s,b_t\neq0\) that are not affected by the higher-order terms in the power series expansions for \(A(r),\; B(r)\).
\\
Then, for the EM to be equal to zero, these coefficients need to vanish. This provides a necessary condition for \((s,t)\) to be the consistent solutions to the EM.
\\
On dimensional grounds, one can see that these coefficients are proportional to the parameters of the highest curvature operators. In this case, this means that the possible families will be determined by the GS counterterm.
\\
These lowest coefficients read, 
   \begin{align}
		   &a_s^4 b_t^6 \Bigg[(t-2)^2 \omega  (s-t+2)^2 \left(30 s^2+s (t+52)-t^2+4 t+20\right)\Bigg]=0,\label{c.1.1.1}\\
		   &a_s^3 b_t^6  \Bigg[(t-2)^3 \omega  (s-t+2)^2 (5 s+t+4)\Bigg]=0.\label{c.2.1.1}
   \end{align}
   Solving this pair of conditions for \((s>0,t)\) on \(\mathbb{Z}\) yields,
   \begin{equation}
	   \bullet \mbox{If } \omega\neq0\Rightarrow \left\lbrace\begin{array}{c} (s,2) \\ (s,2+s) \end{array}\right.\qquad\mbox{for }s\in\mathbb{N}^+.\label{con.2.2}
   \end{equation}
However, these necessary conditions need to be compatible with the terms in the EM corresponding to Einstein's equations. It turns out that the only consistent families are, 
\begin{equation}
   \bullet  (s,t)=(0,0), \qquad
   \bullet (s,t)=(2,2).
\end{equation}
This power series analysis shows that the GS counterterm {\em does not admit an SST}.

   %%%%%%%%%%%%%%%
\subsubsection{(0,0) family.}
In this case, we find three different families, corresponding to three different expansions, 
\begin{itemize}
   \item First we recover Minkowski, that is, up to the considered order, 
   \begin{align}
	   &A(r)=1+\mathcal{O}(r^{10}),\\
	   &B(r)=1+\mathcal{O}(r^{10}).
   \end{align}
   \item Second we find a solution with two free parameters \(\{a_9,b_9\}\),
   \begin{align}
	   &A(r)=1+\frac{r^4}{84 \kappa ^4 \omega }+\frac{1187}{8742384 \kappa ^8 \omega ^2} r^8+a_9r^9+\mathcal{O}(r^{10}),\\
	   &B(r)=1+\frac{r^4}{168 \kappa ^4 \omega }+\frac{8479}{209817216 \kappa ^8 \omega ^2} r^8+b_9r^9+\mathcal{O}(r^{10}).
   \end{align}
   \item And lastly the two-parameter \(\{a_9,b_2\}\) family,
   \begin{align}
   &A(r)=1+b_2r^2\pm \frac{\sqrt{b_2}}{\sqrt{15} \kappa ^2 \sqrt{\omega }}r^3+\left\lbrace\frac{3 b_2^2}{5}+\frac{1}{72 \kappa ^4 \omega }\right\rbrace r^4\pm\nonumber\\
   &\pm\frac{139392 b_2^2 \kappa ^4 \omega -25}{90720 \sqrt{15} \sqrt{b_2} \kappa ^6 \omega ^{3/2}}r^5+\left\lbrace\frac{5}{1161216 b_2 \kappa ^8 \omega^2}+\frac{353 b_2}{4320 \kappa ^4 \omega }+\frac{2 b_2^3}{7}\right\rbrace r^6\mp\nonumber\\
   &\mp\frac{r^7\,\kappa^{-10}}{2^{11} 63^3 5^2 \sqrt{15 b_2^{3}\omega ^{5}}  }\Bigg\{4625-10959058944 b_2^4 \kappa ^8 \omega ^2-91727040 b_2^2 \kappa ^4 \omega \Bigg\}+\nonumber\\
   &+\Bigg\{\frac{127}{4514807808 b_2^2 \kappa ^{12} \omega ^3}+\frac{82319 b_2^2}{567000 \kappa ^4 \omega }+\frac{3 b_2^4}{25}+\frac{46483}{293932800 \kappa ^8 \omega ^2}\Bigg\}r^8\pm\nonumber\\
   &+\Bigg\{\frac{555865 \sqrt{\frac{5}{3}}}{90849972584448 b_2^{5/2} \kappa ^{14} \omega ^{7/2}}+\frac{1318013}{236588470272 \sqrt{15} \sqrt{b_2} \kappa ^{10} \omega ^{5/2}}\mp 9 b_9-\nonumber\\
   &-\frac{19986884 b_2^{7/2}}{3301375 \sqrt{15} \kappa ^2 \sqrt{\omega }}-\frac{8988599813 b_2^{3/2}}{20537193600 \sqrt{15} \kappa ^6 \omega ^{3/2}}\Bigg\}r^9+\mathcal{O}(r^{10})
   \end{align}
   and
   \begin{align}
   &B(r)=1+b_2r^2\pm\frac{2 \sqrt{b_2}}{3 \sqrt{15} \kappa ^2 \sqrt{\omega }}r^3+\left\lbrace\frac{3 b_2^2}{5}+\frac{1}{144 \kappa ^4 \omega }\right\rbrace r^4\pm\nonumber\\
   &\pm\frac{224064 b_2^2 \kappa ^4 \omega -25}{226800 \sqrt{15} \sqrt{b_2} \kappa ^6 \omega ^{3/2}}r^5+\left\lbrace\frac{5}{3483648 b_2 \kappa ^8 \omega ^2}+\frac{3461 b_2}{90720 \kappa ^4 \omega }+\frac{2 b_2^3}{7}\right\rbrace r^6 \mp\nonumber\\
   &\mp\frac{r^7\,\kappa^{-10}}{12^6 35^2 \sqrt{15 b_2^{3}\omega ^{5}} }\Bigg\{4625-5005504512 b_2^4 \kappa ^8 \omega ^2-116826240 b_2^2 \kappa ^4 \omega \Bigg\}+\nonumber\\
   &+\Bigg\{\frac{127}{18059231232 b_2^2 \kappa ^{12} \omega ^3}+\frac{683 b_2^2}{10500 \kappa ^4 \omega }+\frac{3 b_2^4}{25}+\frac{110771}{2351462400 \kappa ^8 \omega ^2}\Bigg\}r^8+\nonumber\\
   &+b_9r^9+\mathcal{O}(r^{10})
\end{align}
   From these power series solutions, one can see that for the small \(r\) expansion the GS counterterm determines the first coefficients. This results in more freedom for the first coefficients which are only subject to the condition
   \begin{equation}
   G(A,A',B,B',B'')=0,
   \end{equation}
   which results in an increase in the number of free parameters for the first few orders. Once the terms corresponding to the Einstein-Hilbert action enter the EM, the number of free parameters ceases to increase.
\end{itemize}
\subsubsection{(2,2) family.}

The expansion for the $(2,2)$ family reads, 
\begin{align}
   &A(r)=a_2r^2+a_3 r^3+a_4 r^4+\mathcal{O}(r^5),\label{22.1}\\
   &\frac{B(r)}{b_2}=r^2+b_3 r^3+\frac{1}{4} \left(\frac{a_3 b_3}{a_2}+4 a_2+b_3^2\right) r^4 +b_5r^5+\mathcal{O}(r^6).\label{22.2}
\end{align}  
To this order, it has five free parameters \((a_2,a_3,a_4,b_3,b_5)\).

This solution was checked up to \(\mathcal{O}(r^9)\). 
But the corresponding constraints on \(b_6(\frac{1}{2\kappa^2},\omega),b_5(\frac{1}{2\kappa^2},\omega)\) and \(b_7(\frac{1}{2\kappa^2},\omega)\) are too lengthy and do not give too much information on the solutions to be explicitly present. 
Note the present terms do not depend on \(1/2\kappa^2\) and as such this expansion will correspond to any possible \((2,2)\) solution of the GS equation, as we checked in section \ref{sec:2.3}.
\subsubsection{On the \((s,s+2),\,(s,2)\) families.}
Aside from these two families, we checked, the rest of the cases in \eqref{con.2.2}
\begin{itemize}
   \item  $(n,2)$  with \(n\in \mathbb{N}/\{2\}\). Requires \(\frac{1}{2\kappa^2}\) to vanish at orders that increase with the value of \(n\). 

The fact that the \((2,2)\) solution holds at orders where the \((3,2)\) family is inconsistent seems to support the idea that \((2,2)\) is a robust solution.

\item The $(n,2\,+n)$ family; with \(n\in \mathbb{N}\), was tested for values \(n=\{0,1,2,3\}\) and again required \(\frac{1}{2\kappa^2}\to 0\).
\end{itemize}
Thus, we can see that in this case, the only solutions that are consistent with the presence of the Einstein-Hilbert action are the  \((0,0)\) and \((2,2)\) families. This last one has been proposed \cite{HoldomP} as a natural end product of compact objects. 

\subsubsection{Singularities at the origin.}
Spacetime singularities \cite{Clarke1994} are a subtle topic. SST is known to have a curvature singularity at \(r=0\).

This is usually seen by evaluating the contraction,
\begin{equation}
   \lim_{r\to 0}\tensor{R}{_\alpha_\beta_\gamma_\delta}\tensor{R}{^\alpha^\beta^\gamma^\delta}=\infty.
\end{equation}
For metric \eqref{metans} this {\em Kretschmann invariant} is given by; 
\begin{align}
   &\tensor{R}{_\alpha_\beta_\gamma_\delta}\tensor{R}{^\alpha^\beta^\gamma^\delta}=\frac{1}{4 r^4 A^4 B^4}\Bigg\{8 r^2 B^4 A'^2+16 (A-1)^2 A^2 B^4+\nonumber\\
   &+r^4 \left(B \left(A' B'-2 A B''\right)+A B'^2\right)^2+8 r^2 A^2 B^2 B'^2\Bigg\},
\end{align}
then, the two families \((0,0)\) and \((2,2)\) satisfy:
\begin{itemize}
   \item If \(A(r)=1+a_2r^2+\cdots\) and \(B(r)=1+b_2r^2+\cdots\), then 
   \begin{equation}
	   \lim_{r\to 0}\tensor{R}{_\alpha_\beta_\gamma_\delta}\tensor{R}{^\alpha^\beta^\gamma^\delta}=12 \left(a_2^2+b_2^2\right) \quad\mbox{finite}.\label{sing1}
   \end{equation} 
   \item  If \(A(r)=a_2r^2+a_4r^4+\cdots\) and \(B(r)=b_2r^2+b_4r^4\cdots\), then 
   \begin{equation}
	   \lim_{r\to 0}\tensor{R}{_\alpha_\beta_\gamma_\delta}\tensor{R}{^\alpha^\beta^\gamma^\delta}\propto \frac{1}{r^8}\to \infty .\label{sing2}
   \end{equation}
\end{itemize}
From \eqref{sing2} we can affirm that the \((2,2)\) solution has a curvature singularity at the origin. Nevertheless, eq. \eqref{sing1} shows no singularity. Thus at least naively it seems that the \((0,0)\) family is regular at the origin. This was further checked by considering other curvature scalars, \(R,\,R^2,\,(\tensor{R}{_\m_\n})^2,\cdots\) since we know from section \ref{sec:2.4} that the \((0,0)\) family cannot be Ricci flat.
%%%%%%%%%%%%%%%%%%%%%%%%%%%%%%%%%%%%%%%%%%%%%%%%%%%%%%%%%%%%%%%%%%%%%%%%%%%%%%%%%%%%%%%%%%%%%%%%%%%%%%%%%%%%%%%%%%%%%%%%%%%%
\subsection{Asymptotic Power Series Solutions.}\label{sec:3.2}
%%%%%%%%%%%%%%%%%%%%%%%%%%%%%%%%%%%%%%%%%%%%%%%%%%%%%%%%%%%%%%%%%%%%%%%%%%%%%%%%%%%%%%%%%%%%%%%%%%%%%%%%%%%%%%%%%%%%%%%%%%
In this section, we consider power series solutions that are asymptotically flat at large radial values. Since we are considering a particular kind of metric there is no need to employ the full Penrose diagram formalism as the more pedestrian and intuitive notions of asymptotic flatness apply here.

To find asymptotically flat solutions to the EM we will assume they admit a power series expansion of the form,
\begin{align}
   A(r)&=1+\sum_n a_nr^{-n},\\
   B(r)&=1+\sum_n b_nr^{-n},
\end{align}
where we do not admit positive powers as they would, for large radial values dominate and take us arbitrarily away from a Minkowski spacetime. 
Solutions of the form above will always ensure we are in a Minkowski spacetime in ordinary spherical coordinates up to \(\mathcal{O}(1/\mathcal{R})\) corrections i.e., 
\begin{equation}
   \tensor{g}{_\mu_\nu}=\tensor{\eta}{_\mu_\nu}+\tensor{h}{_\mu_\nu}, \qquad\mbox{with the entries of}\;\;\tensor{h}{_\mu_\nu}\sim \mathcal{O}(1/\mathcal{R}).
\end{equation}
To find solutions we consider the change of coordinates, 
\begin{equation}
   z:=\frac{1}{r},
\end{equation}
for which the metric \eqref{metric} becomes, 
\begin{equation}
   ds^2=B(z)dt^2-\frac{A(z)}{z^4}dz^2-\frac{1}{z^2}d\Omega_2^2,
\end{equation}
with functions, 
\begin{align}
   A(z)&=1+\sum_n a_nz^{n},\\
   B(z)&=1+\sum_n b_nz^{n}.
\end{align}
Then, using the two independent EM,
\begin{align}
   H_{tt}&=0,\\
   H_{zz}&=0.
\end{align}
Solving the first six orders of the EM such that, 
\begin{align}
   H_{tt}&=0+\mathcal{O}(z^9),\\
   H_{zz}&=0+\mathcal{O}(z^7).
\end{align}
We find the solution expressed again in terms of \(1/r\),
\begin{align}
   A(r)&=1+\frac{a_1}{r}+\frac{a_1^2}{r^2}+\frac{a_1^3}{r^3}+\frac{a_1^4}{r^4}+\frac{a_1^5}{r^5}+\frac{a_1^6-36 a_1^2 \kappa ^4 \omega}{r^6}+\nonumber\\
	   &+\frac{-492 a_1^4 \kappa ^4 \omega +13 a_1^8+7 a_8}{20 a_1}\frac{1}{r^7}+\frac{a_8}{r^8}+\mathcal{O}(\frac{1}{r^9}),\label{70}\\
   B(r)&=1-\frac{a_1}{r}+\frac{b_4}{r^4}-\frac{a_1b_4}{r^5}+\frac{b_6}{r^6}-\frac{-20 a_1^2 b_6+36 a_1^4 \kappa ^4 \omega +a_1^8-a_8}{20 a_1}\frac{1}{  r^7}+\nonumber\\
	   &+\frac{b_8}{r^8}+\mathcal{O}(\frac{1}{r^9}).\label{71}
\end{align}
In the expansions above one can recognize the first terms corresponding to the SST expansion for large \(r\) values,
\begin{align}
   A^{\mbox{\tiny Sch.}}(r)&=1+\sum_{n=1}^{\infty}\left(\frac{a_1}{r}\right)^n=\left(1-\frac{a_1}{r}\right)^{-1},\\
   B^{\mbox{\tiny Sch.}}(r)&=1-\frac{a_1}{r}.
\end{align}
Contrary to the results in section \ref{sec:3.1} now we see that the first orders in the EM are controlled by the Einstein-Hilbert EM. 
On the other hand, the corrections depending on the GS counterterm appear precisely at order \( r^{-6}\) onwards. 
These are the correction to the SST at large distances induced by the GS counterterm.

%%%%%%%%%%%%%%%%%%%%%%%%%%%%%%%%%%%%%%%%%%%%%%%%%%%%%%%%%%%%%%%%%%%%%%%%%%%%%%%%%%%%%%%%%%%%%%%%%%%%%%%%%%%%%%%%%%%%%%%%%%%%%%
\section{Structural stability of Schwarzschild's Event Horizon.}\label{sec:4}
%%%%%%%%%%%%%%%%%%%%%%%%%%%%%%%%%%%%%%%%%%%%%%%%%%%%%%%%%%%%%%%%% %%%%%%%%%%%%%%%%%%%%%%%%%%%%%%%%%%%%%%%%%%%%%%%%%%%%%%%%%%%

It is well-known that, while the Einstein-Hilbert action admits the Schwarzschild metric as a solution, this is not true anymore when the GS operator is included, which does {\em not} admit SST. This point we have repeated throughout the entire paper was first shown by  Deser and Tekin \cite{Deser}. 
This section aims to ascertain where an initial SST is driven in the presence of the GS counterterm.

Let us first remind the reader of some generalities. When a static, spherically symmetric metric is perturbed the resulting spacetime can be quite complicated. Using the notation of \cite{Chandrasekhar} the resulting spacetime can be of the form
\be
ds^2=e^{2\n} dt^2-e^{2\psi}\left(d\phi-q_2 dx^2-q_3 dx^3-\omega dt\right)^2-e^{2\m_1} dx_2^2-e^{2\m_3} dx_3^2.
\ee
Perturbations leading to non-vanishing values of ($\omega$, $ q_2$ and $q_3$) originate a dragging of inertial frames, and are usually referred to as {\em axial}; whereas those leading to changes in ($\n$, $\m_2$, $\m_3$, $\psi$) do not change the rotation, and are called {\em polar}.

In this section, we are interested in analyzing the static spherically symmetric perturbations to SST when the Einstein-Hilbert action is deformed with the GS term, that is,
\be
S=\int d^4 x \sqrt{|g|} \, \Big\{-\frac{1}{2\kappa^2} \, R+\omega\kappa^2\, W^3\Big\}.
\ee  
In particular, since we have shown in section \ref{sec:2.3} that the perturbed EM cannot accommodate an SST we want to explore whether the perturbations introduced by the GS counterterm could give rise to spacetimes without an Event Horizon. 
Such a study in general is complicated as we have not been able to obtain exact solutions to the full EM. It is for this reason that we will argue in favor of the possibility of the GS counterterm removing the Horizon by performing a perturbative analysis.

First, in section \ref{sec:5.1}, we will consider a polar perturbation of the type
\bea
\label{SH}  &&ds^2=\left[1-\frac{r_s}{r}+  B(r)\right]dt^2-\frac{1}{\left[1-\frac{r_s}{r}+ A(r)\right]} dr^2-r^2d\Omega^2.\nonumber\\
\eea
In this formalism, we will consider the first order corrections in \(\omega\kappa^4\) to the EM, that determine the small corrections to \(g_{tt}\) given by \(B(r)\), and discuss whether such a perturbation can remove the horizon.

%%%%%%%%%%%%%%%%%%%%%%%%%%%%%%%%%%%%%%%%%%%%%%%%%%%%%%%%%%%%%%%%%%%%%%%%%%%%%%%%%%%%%%%%%%%%%%
\subsection{Perturbations to the horizons.}\label{sec:5.1}
%%%%%%%%%%%%%%%%%%%%%%%%%%%%%%%%%%%%%%%%%%%%%%%%%%%%%%%%%%%%%%%%%%%%%%%%%%%%%%%%%%%%%%%%%%%%%%

Since the perturbation in \eqref{SH} does not change the static and spherically symmetric nature of the spacetime, we will have that every event horizon will correspond to the killing horizon for a certain timelike vector \(\mathcal{K}\) \cite{Carter}. 
For the particular form of the metric, we are considering, 
\begin{equation}
   \mathcal{K}:=\partial_t,
\end{equation}
is a killing vector. Then the Lie-derivative of the metric with respect to the vector field \(\mathcal{K}\) satisfies, 
\begin{equation}
   \mathcal{L}_{\mathcal{K}}g=\nabla_\a\mathcal{K}_\b+\nabla_\b\mathcal{K}_\a=0.
\end{equation}
The killing horizon of \(\mathcal{K}\) will correspond to a null hypersurface whose orthogonal vector is \(\mathcal{K}\), i.e.
\begin{equation}
   \mathcal{K}^2=1-\frac{r_s}{r}+ B(r)=0.
\end{equation} 
To see how the perturbation \(B(r)\) introduced by the GS counterterm can affect the horizon structure, we consider the first-order EM for \(B(r)\),
\begin{align}
   r^8 A(r)+(r-r_s) \left(r^8 B'(r)+12 \kappa ^2 r_s^2 (4 r_s-3 r)\right)-r^7 r_s B(r)&=0,\label{a11}\\
   r^8 A'(r)+r^7 A(r)+12 \kappa ^2 r_s^2 (16 r_s-15 r)&=0.\label{a22}
\end{align} 
Solving equations, \eqref{a11} and \eqref{a22} gives, 
\begin{align}
   A(r)&=\frac{c_1}{r}+\frac{32 \kappa ^2 r_s^3}{r^7}-\frac{36 \kappa ^2 r_s^2}{r^6}\label{foa},\\
   B(r)&=-\frac{c_2 r_s}{r}+\frac{c_1}{r}+c_2+\frac{8 \kappa ^2 r_s^3}{r^7}-\frac{12 \kappa ^2 r_s^2}{r^6}\label{fob}.
\end{align}
Setting \(c_2=0\) to ensure asymptotic flatness, gives, 
\begin{equation}
   B(r)=\frac{c_1}{r}+\frac{8 \kappa ^2 r_s^3}{r^7}-\frac{12 \kappa ^2 r_s^2}{r^6}\label{Bfo}.
\end{equation}
With equation \eqref{Bfo} we will first consider small displacements of the horizon and then consider the entire parameter space for \(\omega\kappa^2\) to discuss regions in which the horizon can be removed.
\subsection{Small deviations.}
Here we will consider the Killing horizon corresponding to the timelike Killing vector of the metric. If we consider \(r=r_0\) to be the value at which 
\begin{equation}
   g_{tt}(r_0)\big\vert_{\omega=0}=0.
\end{equation}
Then \(r_0=r_s\). If we consider that when \(\omega\neq0\) the new killing horizon is at, 
\begin{equation}
   r_*=r_s+\omega\kappa^2\rho+\cdots.
\end{equation}
We have that \footnote{The right physics is recovered for $c_1=0$ so that no GS term yields back the original position of the horizon.} 
\begin{equation}
   \rho=\frac{4 \kappa ^2}{r_s^3}-c_1,
\end{equation}
and then the position of the horizons is given by, 
\begin{equation}
   r_*:=r_s+\kappa ^4 \omega  \left(\frac{4 }{r_s^3}-c_1\right).
\end{equation}
In this particular case we can check that this value of \(r_*\) makes the first order in \(g^{rr}\) vanish.
%%%%%%%%%%%%%%%%%%%%%%%%%%%%%%%%%%%%%%%%%%%%%%%%%%%%%%%%%%%%%%%%%%%%%%%%%%%%%%%%%%%%%%%%%%%%%%
\subsection{Thermodynamics.}
%%%%%%%%%%%%%%%%%%%%%%%%%%%%%%%%%%%%%%%%%%%%%%%%%%%%%%%%%%%%%%%%%%%%%%%%%%%%%%%%%%%%%%%%%%%%%%
For arbitrarily small perturbations the only effect of the GS counterterm is to so-slightly shift the location of the horizon. 
If we assume no further change takes place in the causal structure of the spacetime, we can in principle calculate the first-order changes that the temperature and entropy associated with the Event Horizon suffer under the GS counterterm. 
Changes in the thermodynamical variables coming from interaction with classical matter (dirty Black Holes) were studied in \cite{Visser1992}, while an analysis for general third-order curvature perturbations was presented in \cite{LuWise}. 
Here we follow the latter to discuss the changes to first order in the perturbative parameter $\omega \kappa^2$. 

We note there are some instances in higher derivative theories of gravity in which the changes in the first Black Hole thermodynamical law \cite{Bekenstein1973} can be exactly calculated \cite{Bueno2016}. 
However, these results depend crucially on the fact that $A(r)=1/{B(r)}$ which allows expressing the series near the horizon in terms of the surface gravity.
We will restrain to a qualitative study of the lowest order and leave the more technical details for elsewhere. 

In order to study the change in temperature, we expand the near-horizon geometry. 

This leads to the known Rindler-like behavior near the horizon. 
Now we analytically continue the geometry to complex coordinates. 
Then, we use the euclidean time periodicity. 

We have that, in order to avoid conical defects at $r=r_\star$, the euclidean time variable has a period whose inverse is related to the temperature. 
To first order, this corresponds to, 
\begin{equation}
   T_H= \frac{\hbar}{4\pi k_B r_s} \left(1+ \frac{4\omega \kappa^4}{r_s^4} \right),\label{t1} 
\end{equation}
which is clearly different from the classical result for the spherically symmetric gravitational collapse \cite{Hawking1975}. 

In order to study the change in the entropy we note that from the asymptotic expansion in \cref{sec:3.2}, the behavior for an observer at infinity has corrections of $\mathcal{O}\left(1/r\right)^{6} $ or higher. 
Therefore the ADM/Komar mass measured by an observer far from the horizon is not changed. Therefore the first law, in this approximation, reads
\begin{equation}
   \frac{dS}{dM}=\frac{1}{T}=\frac{4\pi k_B r_s}{\hbar} \left(1- \frac{4\omega \kappa^4}{r_s^4} \right).
\end{equation}
This naïve calculation shows the first-order change in the thermodynamical quantities associated with the Horizon.
%%%%%%%%%%%%%%%%%%%%%%%%%%%%%%%%%%%%%%%%%%%%%%%%%%%%%%%%%%%%%%%%%%%%%%%%%%%%%%%%%%%%%%%%%%%%%%
\subsection{Removing the horizons.}
%%%%%%%%%%%%%%%%%%%%%%%%%%%%%%%%%%%%%%%%%%%%%%%%%%%%%%%%%%%%%%%%%%%%%%%%%%%%%%%%%%%%%%%%%%%%%%

It is interesting to study the SST modified in the whole region of the parameter space, even though it is unnatural to have adimensional parameters either too big or else too small. We have done that by first taking $c_2=0$ to get asymptotic flatness, and we have found a whole region of values of the constant $c_1$ where the resulting spacetime does not have a horizon. 

\begin{figure}[h!]
   \centering
   \includegraphics[width=0.9\textwidth]{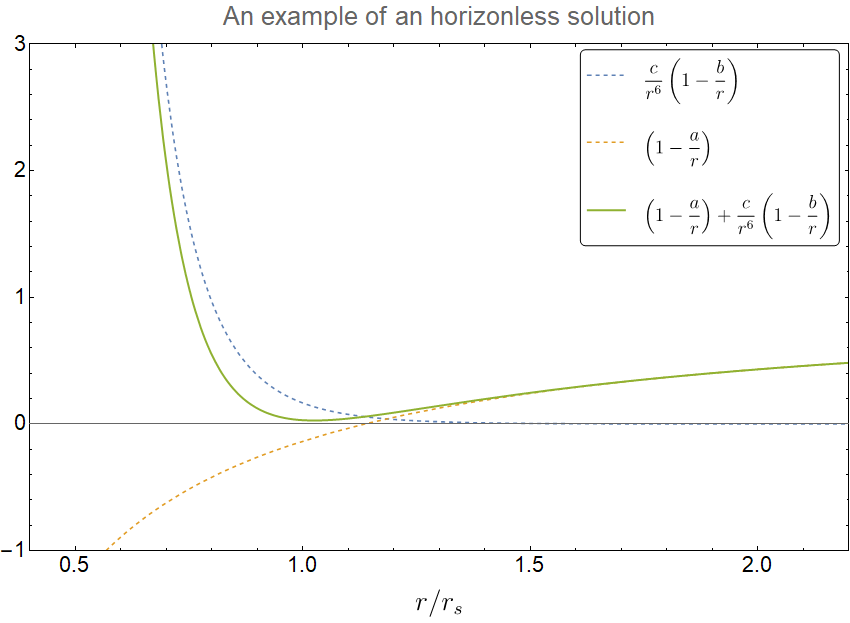}
   \caption{Representation of the \(g_{tt}\) component with linear corrections from the GS counterterm for a choice of the parameters,
   \(a=1.136\),\(\,b=1.898\,\) and \(c=-0.185\), for which there is no horizon.}
   \label{fig:hor1}
\end{figure}

To see that there is a region of the parameter space removing the killing horizon it is useful to rewrite, 
\begin{equation}
   g_{tt}=1-\frac{r_s}{r}+B(r)=1-\frac{a}{r}+\frac{c}{r^6}\left(1-\frac{b}{r}\right),
\end{equation}
where, 
\begin{align}
   a&:=r_s+c_1\omega\kappa^2,\\
   b&:=\frac{2r_s}{3},\\
   c&:=-12r_s^2\omega\kappa^4.
\end{align}
Since we have three free parameters, \(\{r_s,\,c_1,\,\omega\}\), we can freely set \(\{a,\,b,\,c\}\).
Analytic solutions to the configurations of \(\{a,b,c\}\) depend on seventh-order polynomials but from figure \ref{fig:hor1} it is clear that there is a region of the parameter space corresponding to horizonless solutions.
One can further check that the parametric region corresponding to horizonless modifications has a non-zero measure.

While we believe that this is an interesting result, we have to emphasize that our modified spacetime is only valid to the first order in the parameter $\omega$. Definite conclusions have to wait until an exact solution of the whole system (Einstein-Hilbert plus Goroff-Sagnotti) is found.

%%%%%%%%%%%%%%%%%%%%%%%%%%%%%%%%%%%%%%%%%%%%%%%%%%%%%%%%%%%%%%%%%%%%%%%%%%%%%%%%%%%%%%%%%%%%%%%%%%%%%%%%%%%%%%%%%%%%%

%%%%%%%%%%%%%%%%%%%%%%%%%%%%%%%%%%%%%%%%%%%%%%%%%%%%%%%%%%%%%%%%%%%%%%%%%%%%%%%%%%%%%%%%%%%%%%%%%%%%%%%%%%%%%%%%%%%%%%%%%%%%%%
\section{Conclusions.}
%%%%%%%%%%%%%%%%%%%%%%%%%%%%%%%%%%%%%%%%%%%%%%%%%%%%%%%%%%%%%%%%%%%%%%%%%%%%%%%%%%%%%%%%%%%%%%%%%%%%%%%%%%%%%%%%%%%%%%%%%%%%%
The {\em leitmotiv} of the present paper is that we believe that it is physically important to take into account irrelevant (in the renormalization group sense) operators in the gravitational action which might be induced by quantum corrections.
\par
With that in mind, we have examined in some detail the effect of the GS or Weyl cube operator. We found that many arguments could be extended to lagrangians built out of arbitrary powers of Weyl's tensor. We generalized in particular Deser and Ryzhov's arguments to the non-static case, and found the general solution of the EM with spherical symmetry in the non-conformal situation
\par
Concerning the physically relevant EM (including the Einstein-Hilbert contribution), we were able to prove that there are no Ricci flat solutions, but we did not get any new exact solution. Then we have resorted to formal power series solutions to get more information, and we have found a tantalizing $(2,2)$ solution quite similar to the one analyzed by Holdom and Stelle for a few years.
\par
Finally while determining the structural stability of SST  under those deformations, we have just concluded that when the Einstein-Hilbert lagrangian is deformed with the $W^3$ operator the resulting spacetime completely changes its asymptotic behavior. The horizon gets modified, and in some regions of the parameter space, completely disappears.
\par
On the other hand the result would not change very much if we had included quadratic terms in the action. The reason is that quadratic theories admit the SST as a solution, so they do not upset the stability of the initial solution. Only the Weyl cube deformation does that. In the same spirit, the introduction of a cosmological constant can be done easily and it does not affect the result in any essential way.

\par
Although we do not have an analytical grasp of the $(2,2)$ solution,  which, in a neighborhood of $r=0$ behaves as
\be
ds^2 = \left(r^2+\ldots\right)dt^2-\left(r^2+\ldots\right)\,dr^2-r^2 d\Omega^2.
\ee
we have some evidence that it survives this deformation. This comes about from the Frobenius type of power series expansions, as well as from the linearized approach.
We have already mentioned that this solution has been proposed as an alternative to black holes as the end product of stellar collapse. It also has been postulated as a dark matter candidate \cite{Aydemir}.
\par
Were a detailed analysis of the collapsing process to confirm these conjectures, many of the usual arguments involving black hole physics, most of them relying on the presence of a horizon, such as the weak gravity conjecture and holography in general would lose much of their force. 

%%%%%%%%%%%%%%%%%%%%%%%%%%%%%%%%%%%%%%%%%%%%%%%%%%%%%%%%%%%%%%%%%%%%%%%%%%%%%%%%%%%%%%%%%%%%%%%%%%%%%%%%%%%%%%%%%%%%%%%%%%%%%%
\section{Acknowledgements.}
%%%%%%%%%%%%%%%%%%%%%%%%%%%%%%%%%%%%%%%%%%%%%%%%%%%%%%%%%%%%%%%%%%%%%%%%%%%%%%%%%%%%%%%%%%%%%%%%%%%%%%%%%%%%%%%%%%%%%%%%%%%%%
We thank Ángel Murcia for insightful comments regarding higher-derivative Black Holes and their thermodynamical properties. 
We thank professor Antonio López Maroto for useful comments and pointing us to the existence of references \cite{Dobado1993, LuWise}. 
We acknowledge partial financial support by the Spanish MINECO through the Centro de excelencia Severo Ochoa Program  under Grant CEX2020-001007-S  funded by MCIN/AEI/10.13039/501100011033
All authors acknowledge the European Union's Horizon 2020 research and innovation programme under the Marie Sklodowska-Curie grant agreement No 860881-HIDDeN and also byGrant PID2019-108892RB-I00 funded by MCIN/AEI/ 10.13039/501100011033 and by ``ERDF A way of making Europe''.
\newpage
\appendix
\numberwithin{equation}{section}
%%%%%%%%%%%%%%%%%%%%%%%%%%%%%%%%%%%%%%%%%%%%%%%%%%%%%%%%%%%%%%%%%%%%%%%%%%%%%%%%%%%%%%%%%%%%%%%%%%%%%%%%%%%%%%%%%%%%%%%%%%%%%
\section{Some details of the computations}\label{A}
%%%%%%%%%%%%%%%%%%%%%%%%%%%%%%%%%%%%%%%%%%%%%%%%%%%%%%%%%%%%%%%%%%%%%%%%%%%%%%%%%%%%%%%%%%%%
Let us write here for completeness the full EM omitting the explicit \(r\) dependence. Starting with $H_{tt}$:
\begin{align}
&H_{tt}=\frac{1}{144 r^6 A^7 B^5}\Bigg\{132 r^4 \omega  B^4 A'^4 \left(r B'-2 B\right)^2 -48 A^6 B^6 \left(3 \gamma  r^4-8 \omega \right)-\nonumber\\
&-8 A^7 B^6 \left(-18 \gamma  r^4+4 \omega \right)+3 r^2 A^2 B^2\mathcal{F}_\omega\left(A,B,A',B',A'',B'',A^{(3)},B^{(3)}\right)+\nonumber\\
&+A^4\mathcal{G}_\omega\left(A,B,A',B',A'',B'',A^{(3)},B^{(3)},B^{(4)}\right)-\nonumber\\
&-3 r A^3 B\mathcal{H}_\omega\left(A,B,A',B',A'',B'',A^{(3)},B^{(3)},B^{(4)}\right)+\nonumber\\
&+r^3 \omega  A B^3 A'^2 \left(2 B-r B'\right) \Bigg[2 r B \left(63 r A'' B'+A' \left(213 r B''-97 B'\right)\right)-265 r^2 A' B'^2+\nonumber\\
&+4 B^2 \left(149 A'-63 r A''\right)\Bigg]+6 A^5 B^2\mathcal{K}_\omega\left(A,B,A',B',A'',B'',A^{(3)},B^{(3)},B^{(4)}\right)\Bigg\},
\end{align}
with 
\begin{align}
&\mathcal{F}_\omega\left(A,B,A',B',\cdots\right):=105 r^4 \omega  A'^2 B'^4-2 r^3 \omega  B A' B'^2 \left(27 r A'' B'+A' \left(130 r B''+23 B'\right)\right)+\nonumber\\
&+8 r B^3 \Bigg[14\omega r A'^3 B' -2 r^2 \omega  A''^2 B'+\omega  A'^2 \left(r \left(31 B''-14 r B^{(3)}\right)-14 B'\right)+\nonumber\\
&+r \omega  A' \left(\left(32 A''-2 r A^{(3)}\right) B'-23 r A'' B''\right)\Bigg]+4 r^2 \omega  B^2 \Bigg[r^2 A''^2 B'^2+\nonumber\\
&+A'^2 \left(28 r^2 B''^2-60 B'^2+r B' \left(14 r B^{(3)}+43 B''\right)\right)+\nonumber\\
&+r A' B' \left(23 r A'' B''+\left(r A^{(3)}+13 A''\right) B'\right)\Bigg]-\nonumber\\
&-16 B^4 \left(-r^2 \omega  A''^2+14\omega r A'^3 -20 \omega  A'^2+r \omega  A' \left(18 A''-r A^{(3)}\right)\right),
\end{align}
\begin{align}
&\mathcal{K}_\omega\left(A,B,A',B',\cdots\right):=8 B^4 \left(A' \left(3 \gamma  r^5+2 r \omega \right)-14\omega \right)+\nonumber\\
&+49\omega r^4 B'^4 -16\omega r^4 B^3 B^{(4)} -4 r^3 B B'^2 \left(29\omega r B'' -5 \omega B' \right)+\nonumber\\
&+4 r^2 B^2 \Bigg[9\omega r^2 B''^2 +3 \omega B'^2 +2 r B' \left(6\omega r B^{(3)} -3 \omega B''\right)\Bigg],
\end{align}
\begin{align}
&\mathcal{G}_\omega\left(A,B,A',B',\cdots\right):=-4 r^3 B^3 \Bigg[\omega B'^3 \left(40  -87 r A' \right)+26 r^3 \omega  B''^3+\nonumber\\
&+6 r^2 \omega  B' B'' \left(14 r B^{(3)}+B''\right)+6 r \omega  B'^2 \left(r \left(r B^{(4)}-6 B^{(3)}\right)-36 B''\right)\Bigg]+\nonumber\\
&+24 r^2 B^4 \Bigg[-r B' \left(\omega B'' \left(27 r A' -8  \right)+2 r \omega  \left(r B^{(4)}+8 B^{(3)}\right)\right)-\nonumber\\
&-2\omega B'^2 \left(3 r^2 A'' -3 r  A' +2 \right)+2 r^2 \omega  \Big((r B^{(3)})^2 -5 B''^2+\nonumber\\
&+r \left(r B^{(4)}+2 B^{(3)}\right) B''\Big)\Bigg]+48 r^2 B^5 \Bigg[A' \Bigg(\omega B' +r \Big(6\omega r B^{(3)}-\omega B'' \Big)\Big) +\nonumber\\
&+r \Bigg(2\omega r \left(2 A'' B'' +  B^{(4)}\right)+B' \left(\omega r A^{(3)}-\omega A'' \right)\Bigg)\Bigg]-\nonumber\\
&-32 B^6 \Bigg[3 \alpha  r^5 A^{(3)}+18 \beta  r^5 A^{(3)}+\nonumber\\
&+3 r^3 \omega  A^{(3)}-9 r^2 \omega A'' +24 r \omega A' -10 \omega \Bigg]+121 r^6 \omega  B'^6-\nonumber\\
&-6 r^5 \omega  B B'^4 \left(85 r B''-29 B'\right)+\nonumber\\
&+84 r^4 \omega  B^2 B'^2 \left(7 r^2 B''^2-4 B'^2+2 r B' \left(r B^{(3)}-2 B''\right)\right),
\end{align}
and, lastly
\begin{align}
&\mathcal{H}_\omega\left(A,B,A',B',\cdots\right):=28 r^4 \omega  B B'^3 \left(r A'' B'-A' \left(B'-10 r B''\right)\right)-85 r^5 \omega  A' B'^5-\nonumber\\
&-4 r^3 \omega  B^2 B' \Bigg[A' \left(55 r^2 B''^2-46 B'^2+2 r B' \left(10 r B^{(3)}+B''\right)\right)+\nonumber\\
&+r B' \left(18 r A'' B''+\left(r A^{(3)}+A''\right) B'\right)\Bigg]+\nonumber\\
&+8 r B^4 \Bigg[\omega r A'^2 \left(19 r B'' -5 B' \right)-2 r \omega  \Bigg(2 \left(A''-r A^{(3)}\right) B'+\nonumber\\
&+r \left(r A^{(3)} B''+A'' \left(2 r B^{(3)}-5 B''\right)\right)\Bigg)+A' \Bigg(B' \left(13\omega r^2 A'' +4 \omega \right)-\nonumber\\
&-2 r \omega  \left(2 B''+r \left(r B^{(4)}-8 B^{(3)}\right)\right)\Bigg)\Bigg]+2 r^2 B^3 \Bigg[-57\omega r A'^2 B'^2 +4 \omega  A' \Bigg(13 B'^2+\nonumber\\
&+2 r^2 B'' \left(6 r B^{(3)}+5 B''\right)+B' \left(r^3 B^{(4)}-46 r B''\right)\Bigg)+\nonumber\\
&+4 r \omega  \Bigg(4 r^2 A'' B''^2-10 A'' B'^2+r B' \Big(r A^{(3)} B''+\nonumber\\
&+A'' \left(2 r B^{(3)}+5 B''\right)\Big)\Bigg)\Bigg]+8 B^5 \Bigg[ 29 r \omega A'^2-4 r \omega  \left(r A^{(3)}-3 A''\right)-\nonumber\\
&-2\omega  A' \left(13 r^2 A'' +14  \right)\Bigg].
\end{align}
The component $H_{rr}$ in turn reads:
\newpage
\begin{align}
&H_{rr}=\frac{1}{144 r^6 A^5 B^6}\Bigg\{8 \left(-18 \gamma  r^4+4 \omega \right) A^6 B^6+11  \omega  (rA' \left(r B'-2 B\right) B)^3+\nonumber\\
+&48 (A\,B)^5 \left(\left(3 \gamma  r^4-4 \omega \right) B+r \left(3 \gamma  r^4+2 \omega \right) B'\right)+3 (r B)^2 \omega  A A' \left(r B'-2 B\right)^2 \Bigg[7 r^2 A' B'^2+\nonumber\\
-&2 r \left(r B' A''+A' \left(6 r B''-4 B'\right)\right) B-4 \left(5 A'-r A''\right) B^2\Bigg] +6 A^4 B^6 \Bigg[48 \omega B^2-\nonumber\\
-&16\omega  B{\Bigg( } r^3 B^{(3)}+2r   B'{\Bigg)} -4 r^2 {\Bigg( }\omega B'^2-4\omega r \left( B''+r B^{(3)}\right) B'+r^2\omega  B''^2{\Bigg) } -\nonumber\\
-&12\omega r^3 B^{-2}B'^2 \left(  B'+ r  B''\right) B+7\omega r^4  B^{-2}B'^4\Bigg] +3 r A^2 B\Bigg[8\omega   B^5\left(7r A'^2-8   A'+4 r  A''\right)+\nonumber\\
+&2 r^2 \Bigg(7\omega r  A'^2 B'^2+4 \omega  A' B^3 \left(3 B'^2+r \left(B''-2 r B^{(3)}\right) B'-2 r^2 B''^2\right)+\nonumber\\
+&8 r \omega  A'' \left(B'-r B''\right) B'\Bigg)-8 r B^4\Big[7\omega r B' A'^2-2 \omega  \left(4 B'+r \left(r B^{(3)}-3 B''\right)\right) A'-\nonumber\\
-&2 r \omega   A'' \left(r B''-3 B'\right)\Big]+4 r^3 \omega  B' B^2\Bigg(A' \left(2 r^2 B''^2-5 B'^2+r \left(11 B''+r B^{(3)}\right) B'\right)+\nonumber\\
+&r B' A'' \left(B'+r B''\right)\Bigg)-2 r^4 \omega  B'^3 \left(r B' A''+A'B \left(7 B'+9 r B''\right)\right) B+7 r^5 \omega  A' B'^5 \Bigg] +\nonumber\\
+&A^3 \Bigg(11 \omega  B'^6 r^6-42 \omega  B B'^4 B'' r^6+12 \omega  B^2 B'^2 \left(-8 B'^2+r \left(5 B''+r B^{(3)}\right) B'+3 r^2 B''^2\right) r^4-\nonumber\\
+&4 (r B)^3 \left(\omega\left(16  +9 r A'\right) B'^3+48 r \omega  B'' B'^2-6 r^2 \omega  B'' \left(5 B''+r B^{(3)}\right) B'+2 r^3 \omega  B''^3\right) +\nonumber\\
+&24 B^4 \Bigg(2 \omega  B'' \left(B''+r B^{(3)}\right) r^2-2 rB' \left(\omega\left(5  +r A' \right) B''+2 r \omega  B^{(3)}\right) -\nonumber\\
-&B'^2 \left(\omega A'' r^2+2 \omega A' r-2 \omega \right)\Bigg) r^2+48 r B^5 \Bigg(2\omega A' B''+ B^{(3)} r^2+\nonumber\\
+&B' \left(-3 r \omega  A'+2\omega \left(A'' r^2+1\right)\right)\Bigg)-32\omega B^6 \left(3 A'' r^2-6  A' r+4 \right)\Bigg).
\end{align}
where $A^{(n)}$ corresponds to the n-th partial derivative.
%%%%%%%%%%%%%%%%%%%%%%%%%%%%%%%%%%%%%%%%%%%%%%%%%%%%%%%%%%%%%%%%%%%%%%%%%%%%%%%%%%%%%%%%%%%%%%%%%%%%%%%%%%%%%%%%%%%%%%%%%%%
\section{Details of the explicit derivation when \(p=3\) (GS).}\label{sec:B}
%%%%%%%%%%%%%%%%%%%%%%%%%%%%%%%%%%%%%%%%%%%%%%%%%%%%%%%%%%%%%%%%%%%%%%%%%%%%%%%%%%%%%%%%%%%%%%%%%%%%%%%%%%%%%%%%%%%%%%%%%%%%
In this appendix, we check explicitly the property derived for general \(p\) in section \ref{sec:2.1} for the case \(p=3\). We also present the diff-invariant EM whose trace is used in section \ref{sec:2.4}.

The above general result can be explicitly checked for the case of the GS counterterm.

\textbf{Claim:}
For the GS counterterm, all non-degenerate, spherically symmetric, and static solutions to the EM are obtained by solving for, 
\begin{equation}
   G(A,A',\dot{A},\ddot{A},B,B',B'',\dot{B})=0.
\end{equation}

\textbf{Proof:}
Using Lanczos' identity, for the four-dimensional Weyl tensor \cite{Lanczos},
\be 
W_{\m\a\b\l}W_{\n}^{~\a\b\l} =\frac{1}{4}g_{\m\n}W^2.
\ee
The EM for the GS correspond to 
\begin{align}
V_{\m\n}&\equiv {1\over \sqrt{|g|}}{\d S^{(GS)}\over \d g^{\m\n}}={1 \over 2} W^3 g_{\m\n} +{1\over 4}\left(R g_{\m\n}- R_{\m\n}\right) W^2 -{1\over 2} \left(\nabla_\m\nabla_\n -\Box g_{\m\n}\right) W^2-\nonumber\\
&-3 R^{\a\b} (W^2)_{\m\a\n\b}-3 W_\m\,^{\a\b\l} (W^2)_{\n\a\b\l} -6 \nabla^\a\nabla^\b (W^2)_{\m\a\n\b},\label{EM}
\end{align}
where
\be
(W^2)_{\m\n\r\s}\equiv W_{\m\n\a\b} W^{\a\b}\,_{\r\s}.
\ee
Note that \(\tensor{V}{_\mu_\nu}\) is a primary operator of dimension 4, that is, under a conformal transformation,
\begin{equation}
   \tensor{g}{_\mu_\nu}\rightarrow\Omega(x)^2\tensor{g}{_\mu_\nu}\Rightarrow\tensor{V}{_\mu_\nu}\rightarrow\Omega(x)^{-4}\tensor{V}{_\mu_\nu}.
\end{equation}

\begin{itemize}
   \item \(\Rightarrow\)
   \\
   From the form of the Weyl tensor for a metric \eqref{metric2}, it is clear that, 
   \begin{equation}
	   G=0\Rightarrow\tensor{W}{^\a_\b_\g_\d}\quad \mbox{and}\quad W^2=0.
   \end{equation}
   It is clear then that \(A(r),\,B(r)\) satisfying \(G=0\) will be solutions to equations \eqref{EM}.
   \item \(\Leftarrow\)
   \\
   To see that this is a sufficient condition for the vacuum EM to vanish we trace the EM, 
   \begin{equation}
	   \tensor{V}{_\mu_\nu}=0\Rightarrow \tensor{g}{^\mu^\nu}  \tensor{V}{_\mu_\nu}=\frac{n-6}{2}W^3\propto{G^3}=0.\label{41}
   \end{equation}
Then we have from \eqref{41} that \(G=0\) is a necessary and sufficient\footnote{At least when the metric is non singular, \(A(r),B(r)\neq0\). } condition for \(A(r),\,B(r)\) to be solutions to the EM \eqref{EM} in vacuum.

\end{itemize}
\newpage
%%%%%%%%%%%%%%%%%%%%%%%%

\end{document}